\documentclass[]{spie}  %>>> use for US letter paper
\usepackage{amsmath,amsfonts,amssymb}
\usepackage{graphicx}
\usepackage[colorlinks=true, allcolors=blue]{hyperref}

\title{Optical turbulence forecast over short timescales using machine learning techniques}

\author[a]{Turchi, A.}
\author[a]{Masciadri, E.}
\author[a]{Fini, L.}
\affil[a]{INAF-Osservatorio Astrofisico di Arcetri, L.go Enrico Fermi 5, Firenze, Italy}

\authorinfo{Send correspondence to Alessio Turchi - e-mail: alessio.turchi@inaf.it}

\begin{document} 
\maketitle

\begin{abstract}
Forecast of optical turbulence and atmospheric parameters relevant for ground-based astronomy is becoming an important goal for telescope planning and AO instruments optimization in several major telescope. Such detailed and accurate forecast is typically performed with numerical atmospheric models. Recently short-term forecasts (a few hours in advance) are also being provided (ALTA project) using a technique based on an autoregression approach, as part of a strategy that aims to increase the forecast accuracy. It has been proved that such a technique is able to achieve unprecedented performances so far. Such short-term predictions make use of the numerical model forecast and real-time observations. In recent years machine learning (ML) techniques also started to be used to provide an atmospheric and turbulence forecast. Preliminary results indicate however an accuracy not really competitive with respect to the autoregressive method or even prediction by persistence. This technique might be applicable joint to atmospheric model. It is therefore interesting to investigate the main features of their performances and characteristics (also because there is a great number of algorithms potentially accessible) to understand if results achieved so far can be further improved using ML. In this study we focus on a purely machine learning application to short term forecast (1-2 hours) of astroclimatic and other atmospheric parameters above VLT.
\end{abstract}

\keywords{simulations,machine learning,forecast,optical turbulence,atmosphere,adaptive optics}

\section{INTRODUCTION}
\label{sec:intro} 
Accurate forecast of specific atmospheric parameters, including Optical Turbulence (OT), is becoming a fundamental tool for ground-based astronomy, with the complexity of the instruments growing larger and larger toward the Extremely Large Telescope (ELT) era. OT is one of the main limiting factors for the achievement of high angular resolution from ground-based observations, though in general the atmospheric conditions impact on all the observations performed on top-class telescopes (e.g.  Precipitable Water Vapor (PWV) limiting infrared observations [\citenum{turchi2020}]). To overcome OT limitations, Adaptive Optics (A0) instruments were developed in the last decades, however also their performance depends on the atmospheric and OT conditions, and peak correction can be achieved only in specific and relatively rare cases of perfect weather, which depends not only on the seeing ($\epsilon$) OT parameter, but also on the isoplanatic angle ($\theta_0$) and the wavefront coherence time ($\tau_0$) [\citenum{rabien2019}, \citenum{neichel2014}, \citenum{pedichini2016}]. The large adoption of Wide Field AO (WFAO) adaptive optics systems, and their growing complexity, means that in the future the atmospheric conditions will have an even more deep impact on the scientific output of a top-class telescope, and their ability to fully utilize their AO capabilities to approach diffraction-limited performance. An atmospheric and OT forecast is thus critical in optimizing the schedule of a telescope by matching observations with the required atmospheric condition that will allow an optimal scientific output [\citenum{milli2019}].
The ALTA Center project\footnote{\href{http://alta.arcetri.inaf.it/legend.php\#shortime}{http://alta.arcetri.inaf.it}}, i.e. the OT and atmospheric forecast tool supporting the Large Binocular Telescope (LBT), has seen heavy development in recent years [\citenum{turchi2017,masciadri2020}]. A similar forecast tool is under development for ESO's Very Large Telescope (VLT). Preliminary results on forecast performances are summarized in [\citenum{masciadri2022}].

These tools are using the Astro-Meso-NH mesoscale atmospheric model [\citenum{masciadri1999}] that simulates the whole atmosphere in a limited but high resolution region centered on the telescope, and provide excellent results in terms of accuracy (see [\citenum{masciadri2020}] for a short summary), and typically provide long-term forecasts which are made available few hours before sunset and covering the whole night. 

In recent years however new forecast techniques are becoming available, which use autoregression (AR) methods to combine the long-term forecast of mesoscale models and the telemetry data coming from the telescope sensors, and are able to provide short-term forecasts over a window of 1-4 hours, regularly updated during the night of observations [\citenum{masciadri2020}].\\
Short-term AR aided forecasts are able to provide a huge gain, in terms of forecast error, on all the parameters interested by this kind of prediction [\citenum{masciadri2020}], though being limited to few hours in the future, and allow the telescope operations to think about a different kind of planning. A first implementation of such strategy is currently present in the ALTA project.\\

Other studies focused on implementing an OT forecast purely based on machine learning (ML), without the input from an atmospheric model (which is based on physics), relying only on the measurements made available from the telescope instruments and monitors [\citenum{milli2019}]. These methods seem to provide a limited accuracy with respect to the previously mentioned AR short-term forecasts, and share the same limitations on the future forecast windows of very few hours. Despite this the implementation of such tools is very preliminary and it's worth studying their performances in order to explore their capabilities. The present paper is dedicated to this latter aspect. Also any knowledge accumulated with these tools could prove instrumental in increasing the performances of the already implemented short-term predictions, which huge benefit for the telescope planning and scientific output.\\

In this contribution we will concentrate our attention on the Random Forest (RF) ML algorithm as it has been already used for atmospheric forecasts [\citenum{milli2019}]. We investigate the feasibility of a forecast of Optical Turbulence (astroclimatic) parameters (seeing, wavefront coherence time ($\tau_0$), Isoplanatic Angle ($\theta_0$) and Ground Layer Fraction (GLF), that is the $C_N^2$ fraction at ground, and atmospheric parameters (Temperature, Relative Humidity, Wind speed and direction), by using only instrument data from the telescope telemetry. This study focuses on the VLT telescope and make use of data obtained from the Ambient Condition Database provided by ESO\footnote{\href{http://archive.eso.org/cms/eso-data/ambient-conditions.html}{http://archive.eso.org/cms/eso-data/ambient-conditions.html}}, which by far is one of the most complete collection of telemetry data with a wide range of measured parameters with different sensors, without any input from the long-term mesoscale model forecast. Specifically, we are interested into characterizing the behaviour of the ML method with respect to different parameters such as the training sample length and a variation of the sampling temporal frequency of dataset. We also test two different applications for 1-hour and 2-hour future forecast, which are the most relevant for telescope real-time applications. The aim is to evaluate the reliability of the ML method itself and pave the way to more complex applications, which may also make use of input parameters coming from an atmospheric forecast model. We are interested in identifying and characterizing the constraints imposed by the ML method. For the sake of simplicity, in this preliminary study we use the RMSE error as the sole indicator for the forecast performance. Once the methods and the optimal input sets are selected, we will focus also on the LBT telescope implementation in future studies. We refer to ESO database for an in-detail explanation of all the parameters treated in this study.\\

\section{Algorithm and input set definition}
\label{sec:ml}
ML saw a huge development in the last decades of XX century and rose to a widespread usage in the first decades of XXI century. The term can be used as a general hat to cover different disciplines from Artificial Intelligence to Neural Networks and Computational Statistics. In general we refer to ML techniques when based on algorithms that make use of heterogeneous data to automatically ``learn'' and build a  ``model'' that is used to produce a desired output, using statistical methods. While a general discussion on the several categories of ML methods is out of the scopes of this paper, in order to perform an atmospheric and OT prediction we are interested in the broad class of Supervised methods, that is algorithms that are trained over pre-given set of inputs and outputs. Among this general category, algorithms can be divided into Classifiers, that is methods that predicts general categories as an output (e.g. bad/good seeing) and Regressors, that instead produce a real number (i.e. the value of the seeing). This paper focuses on the Random Forest (RF) algorithm [\citenum{andy2002}], already used in previous similar studies [\citenum{milli2019}]. The RF algorithm is one of the simplest yet robust methods that can be implemented, and thus is our choice for this initial investigation of the problem. The algorithm used in this study can be used as both Classifier or Regressor, however we will limit to the Regressor case for the sake of simplicity, leaving the study of Classifiers to a future paper.\\

%Specifically, after evaluating several options, such as Support Vector Regression and K-Neighbours methods, we decided to focus on Random Forest (RF) and Multi-Layer-Perceptron (MLP) algorithms, which by far produce the best results among the methods we tested. Future studies will surely expand the study to different methods.\\

Random Forest Regression algorithm [\citenum{andy2002}] is a decision-tree based method that tries to overcome the limitations of its simpler cousins (namely overfitting of the training set) by averaging the result over several trees. In the present study we selected 200 trees for the RF, seeing almost no benefit in using more in terms of statistical variation of the output. We used the scikit-learn implementation of RF [\citenum{pedregosa2011}] in this study.\\
RF operates by training the decision trees over a training set of input data by minimizing the mean squared error between the produced output and the given pre-known output, while the performances are validated on a testing set which is independent from the training one, and where previous knowledge of the output is not given to the algorithm.\\
We focus in this contribution over 1-hour and 2-hour future forecast time scale. We remember that a different training has to be associated to a precise time scale forecast. Example: if one wants to perform forecasts at a 2h time scales, he has to perform a training specific for that time scale. Our RF implementation executes a forecast at each time available in the data (i.e. every 5 min if this is the sampling time), using 1 or 2 hour past measurements as an input, respectively to forecast 1 hour or 2 hours in the future. These past data buffers must have continuous consecutive data points (i.e. no holes) in order to successfully perform the forecast. We test the performance of the method by comparing the RMSE averaged over all the forecasts computed on the test dataset.\\

%Multi-Layer-Perceptron Regression algorithm [\citenum{murtagh1991}] are a generic class of Artificial Neural Networks (ANN), that have a wide range of applications. In general a MLP maps an input layer to an output layer through a variable number of hiddel layers, each made of nodes called ``neurons'', corresponding to a non-linear activation function. though this function, each node is fully connected to all other nodes of the following layer, although with different weights.\\

\begin{figure}
\centering
\includegraphics[width=0.7\textwidth]{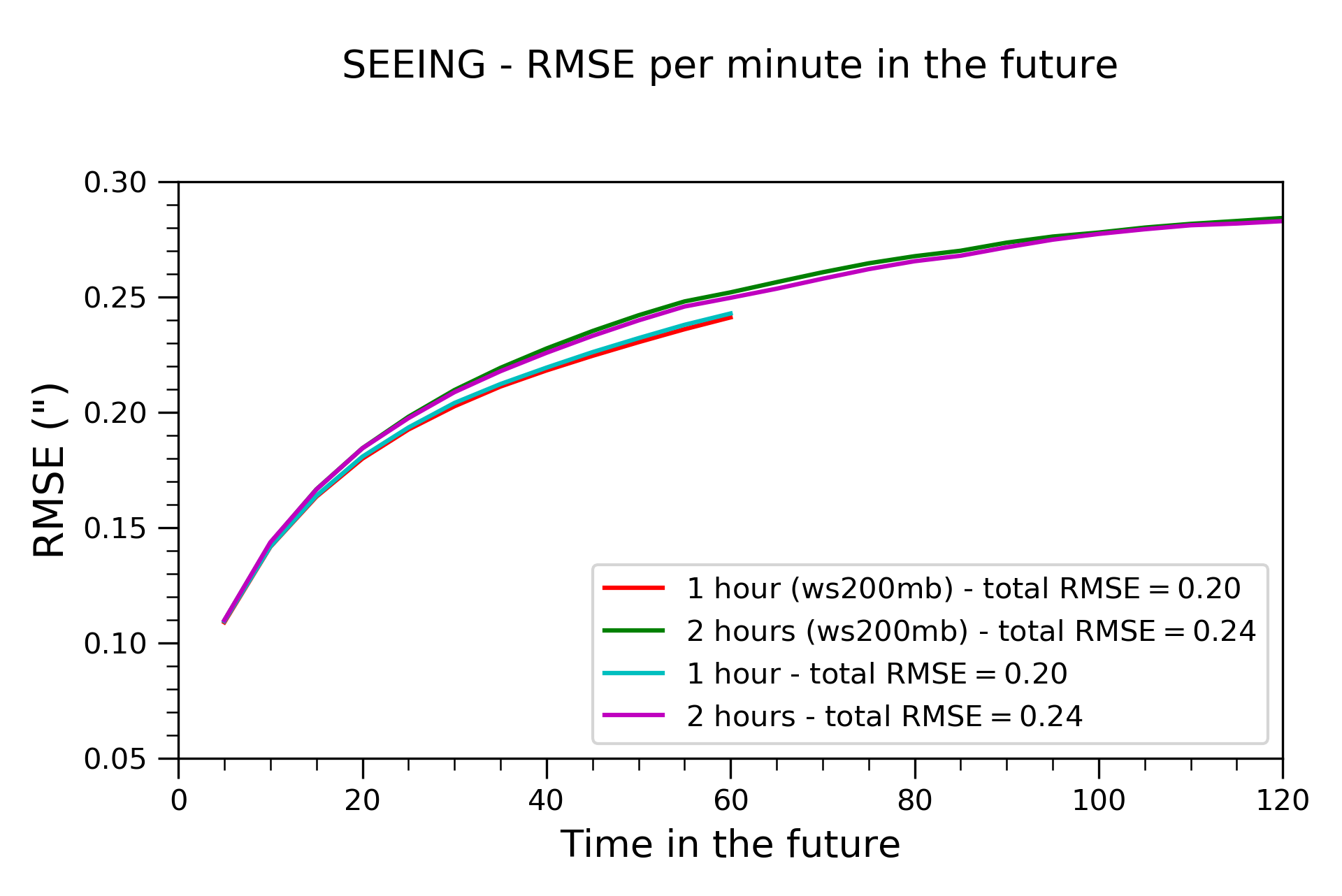}
\caption{Impact of jet-stream level wind speed (at 200mb) on seeing forecasts performed with RF algorithm. Forecasts are performed over 1 or 2 hours in the future, either with or without the 200mb wind speed input parameter, showing negligible differences in terms of RMSE. Time axis is in minutes.}
\label{fig:ws200mb}
\end{figure}

Taking inspiration from Milli, J. et. al. [\citenum{milli2019}], we decided to start by selecting the same set of input parameters. Milli's paper also make use of wind speed measurements at 200mb (jet stream level), however such measurements are not available in ESO database and are technically complex to obtain, so the impact of 200mb wind is evaluated separately. To do so we selected a limited data period from 2018-08-01 to 2019-08-01, with the training set from 2018-08-01 to 2019-02-01 (0.5 years) and Test set from 2019-02-01 to 2019-08-01 (0.5 years), where we managed to obtain 200mb wind speed data from simulations done with a mesoscale model, since this solution has already proven able to reconstruct the vertical profile of the wind with extreme precision [\citenum{masciadri2013,sivo2018}]. In Fig. \ref{fig:ws200mb} we report the results of this test, that show negligible difference in model performances (order $10^-3$), in terms of RMSE, so we decided to remove the 200mb wind speed from the rest of the analysis.\\

\begin{figure}
\centering
\includegraphics[width=0.7\textwidth]{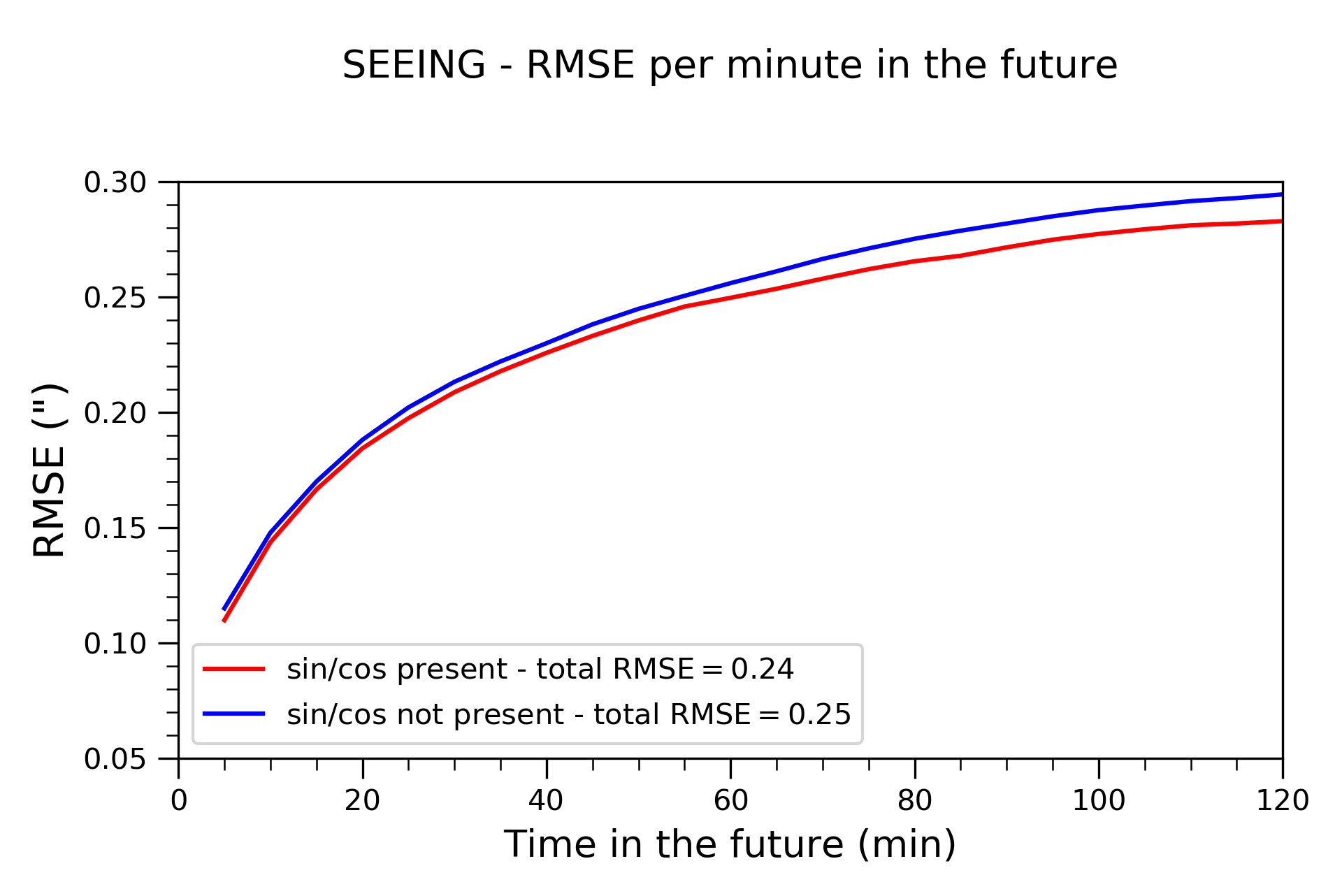}
\caption{Impact of the sin/cos terms on the RF forecast of the seeing over 2 hours in the future. The presence of the terms in the input data proved a small but not negligible effect on the performance.}
\label{fig:sincos}
\end{figure}

Milli's paper also adds a cyclical representation of the day of the year (sin/cos(sin/cos(Day\_of\_Year/365), numbering each day from 1 to 365) and of the hour of the day (sin/cos(Hour\_of\_Day/24), numbering each hour from 1 to 24), which helps the ML algorithm to build correlations with seasonal and daily variations of meteorological parameters and have a little but not negligible effect, as shown in Fig. \ref{fig:sincos}.\\

The full set of input parameters used in this paper to forecast OT parameters (seeing $\epsilon$, isoplanatic angle $\theta_0$, wavefront coherence time $\tau_0$, $C_N^2$ fraction at ground (GLF), is reported in table \ref{tab:defparamastro}.  The input set used to forecast atmospheric parameters (temperature, relative humidity (RH), wind speed at 30m (WS) and Wind direction at 30m (WD), is reported in table \ref{tab:defparamatmo}.\\

\begin{table}[ht]
\caption{Selected input parameters used for OT parameters ($\epsilon$, $\theta_0$, $\tau_0$, GLF)} 
\label{tab:defparamastro}
\begin{center}       
\begin{tabular}{|l|l|} %% this creates two columns
%% |l|l| to left justify each column entry
%% |c|c| to center each column entry
%% use of \rule[]{}{} below opens up each row
\hline
\rule[-1ex]{0pt}{3.5ex}  {\bf Parameter} & {\bf Measurement device}  \\
\hline
\rule[-1ex]{0pt}{3.5ex}  Seeing $\epsilon$ & DIMM   \\
\hline
\rule[-1ex]{0pt}{3.5ex}  Wavefront coherence time $\tau_0$ & MASS-DIMM   \\
\hline
\rule[-1ex]{0pt}{3.5ex}  Isoplanatic angle $\theta_0$ & MASS-DIMM   \\
\hline
\rule[-1ex]{0pt}{3.5ex}  $C_N^2$ fraction at ground (GLF) & MASS-DIMM   \\
\hline
\rule[-1ex]{0pt}{3.5ex}  Air Pressure at ground & VAISALA station \\
\hline
\rule[-1ex]{0pt}{3.5ex}  Air Temperature at 30m & VAISALA station  \\
\hline
\rule[-1ex]{0pt}{3.5ex}  Wind Speed U (horizontal) direction at 20m & VAISALA station  \\
\hline
\rule[-1ex]{0pt}{3.5ex}  Wind Speed V (horizontal) direction at 20m & VAISALA station  \\
\hline
\rule[-1ex]{0pt}{3.5ex}  Wind Speed W (vertical) direction at 20m & VAISALA station  \\
\hline
\rule[-1ex]{0pt}{3.5ex}  sin(Day\_of\_Year/365) &  \\
\hline
\rule[-1ex]{0pt}{3.5ex}  cos(Day\_of\_Year/365) & \\
\hline 
\rule[-1ex]{0pt}{3.5ex}  sin(Hour\_of\_Day/24) & \\
\hline
\rule[-1ex]{0pt}{3.5ex}  cos(Hour\_of\_Day/24) &   \\
\hline
\end{tabular}
\end{center}
\end{table}

\begin{table}[ht]
\caption{Selected input parameters used for atmospheric parameters (temperature, relative humidity, precipitable water vapor, wind speed and direction)} 
\label{tab:defparamatmo}
\begin{center}       
\begin{tabular}{|l|l|} %% this creates two columns
%% |l|l| to left justify each column entry
%% |c|c| to center each column entry
%% use of \rule[]{}{} below opens up each row
\hline
\rule[-1ex]{0pt}{3.5ex}  {\bf Parameter} & {\bf Measurement device}  \\
\hline
\rule[-1ex]{0pt}{3.5ex}  Air Temperature at 30m & VAISALA station   \\
\hline
\rule[-1ex]{0pt}{3.5ex}  Air Temperature at 2m & VAISALA station   \\
\hline
\rule[-1ex]{0pt}{3.5ex}  Air Pressure at ground & VAISALA station   \\
\hline
\rule[-1ex]{0pt}{3.5ex}  Wind Speed at 30m (module) & VAISALA station   \\
\hline
\rule[-1ex]{0pt}{3.5ex}  Relative Humidity at 30m & VAISALA station \\
\hline
\rule[-1ex]{0pt}{3.5ex}  Relative Humidity at 2m & VAISALA station  \\
\hline
\rule[-1ex]{0pt}{3.5ex}  Wind Direction at 30m & VAISALA station  \\
\hline
%\rule[-1ex]{0pt}{3.5ex}  Precipitable Water Vapour (PWV) & LHATPRO  \\
%\hline
\rule[-1ex]{0pt}{3.5ex}  sin(Day\_of\_Year/365) &  \\
\hline
\rule[-1ex]{0pt}{3.5ex}  cos(Day\_of\_Year/365) & \\
\hline 
\rule[-1ex]{0pt}{3.5ex}  sin(Hour\_of\_Day/24) & \\
\hline
\rule[-1ex]{0pt}{3.5ex}  cos(Hour\_of\_Day/24) &   \\
\hline
\end{tabular}
\end{center}
\end{table}

\section{Forecast with Random Forest algorithm}
\label{sec:rf}

The first test that we are interested to perform is to understand how the performance of the RF algorithm changes with the length of the dataset used to train the method. For this purpose we selected data from different training sets defined in tables \ref{tab:trainlenatmo} for atmospheric parameters (left table) and OT parameters (right table). This test is fundamental to understand how much past data is requested to saturate the ML performance.\\
We were forced to select different periods for atmospheric and OT parameters training sets because OT monitor instruments (DIMM and MASS) were updated after April 2016, thus forcing us to use a smaller sample for OT forecasts in order to have homogeneous measurements taken with the same instrument. For this initial test we decided to perform a resampling average over 5 minutes for all the input data.\\

\begin{table}[ht]
\caption{Left:Training set duration for scaling test on atmospheric parameters (temperature, relative humidity, wind speed and wind direction). Test set is 2019/01/01 - 2021/06/01 (2.5 years). Right: Training set duration for scaling test on OT astroclimatic parameters (seeing, $\tau_0$, $\theta_0$, GLF). Test set is 2020/06/30 - 2020/12/31 (0.5 years).} 
\label{tab:trainlenatmo}
\centering
\begin{minipage}{.49\textwidth}
\label{tab:trainlenatmo}
\begin{center}       
\begin{tabular}{|l|l|} %% this creates two columns
%% |l|l| to left justify each column entry
%% |c|c| to center each column entry
%% use of \rule[]{}{} below opens up each row
\hline
\multicolumn{2}{c}{Atmospheric parameters}  \\
\hline
\rule[-1ex]{0pt}{3.5ex}  {\bf Training sample period} & {\bf Total time}  \\
\hline
\rule[-1ex]{0pt}{3.5ex}  2006/01/01 - 2015/01/01 & 9 years   \\
\hline
\rule[-1ex]{0pt}{3.5ex}  2015/01/01 - 2019/01/01 & 4 years   \\
\hline
\rule[-1ex]{0pt}{3.5ex}  2017/01/01 - 2019/01/01 & 2 years   \\
\hline
\rule[-1ex]{0pt}{3.5ex}  2018/01/01 - 2019/01/01 & 1 year \\
\hline
\rule[-1ex]{0pt}{3.5ex}  2018/06/01 - 2019/01/01 & 7 months  \\
\hline
\rule[-1ex]{0pt}{3.5ex}  2018/11/01 - 2019/01/01 & 2 months  \\
\hline
\end{tabular}
\end{center}
\end{minipage}
\hfill
\begin{minipage}{.49\textwidth}
\begin{center}       
\begin{tabular}{|l|l|} %% this creates two columns
%% |l|l| to left justify each column entry
%% |c|c| to center each column entry
%% use of \rule[]{}{} below opens up each row
\hline
\multicolumn{2}{c}{OT parameters}  \\
\hline
\rule[-1ex]{0pt}{3.5ex}  {\bf Training sample period} & {\bf Total time}  \\
\hline
\rule[-1ex]{0pt}{3.5ex}  2016/04/01 - 2020/06/30 & 4.2 years   \\
\hline
\rule[-1ex]{0pt}{3.5ex}  2018/01/01 - 2020/06/30 & 2.5 years   \\
\hline
\rule[-1ex]{0pt}{3.5ex}  2019/01/01 - 2020/06/30 & 1.5 years   \\
\hline
\rule[-1ex]{0pt}{3.5ex}  2019/06/01 - 2020/06/30 & 1 year \\
\hline
\rule[-1ex]{0pt}{3.5ex}  2020/01/01 - 2020/06/30 & 6 months  \\
\hline
\rule[-1ex]{0pt}{3.5ex}  2020/03/01 - 2020/06/30 & 3 months  \\
\hline
\end{tabular}
\end{center}
\end{minipage}
\end{table}

\begin{figure}
\centering
\includegraphics[width=0.85\textwidth]{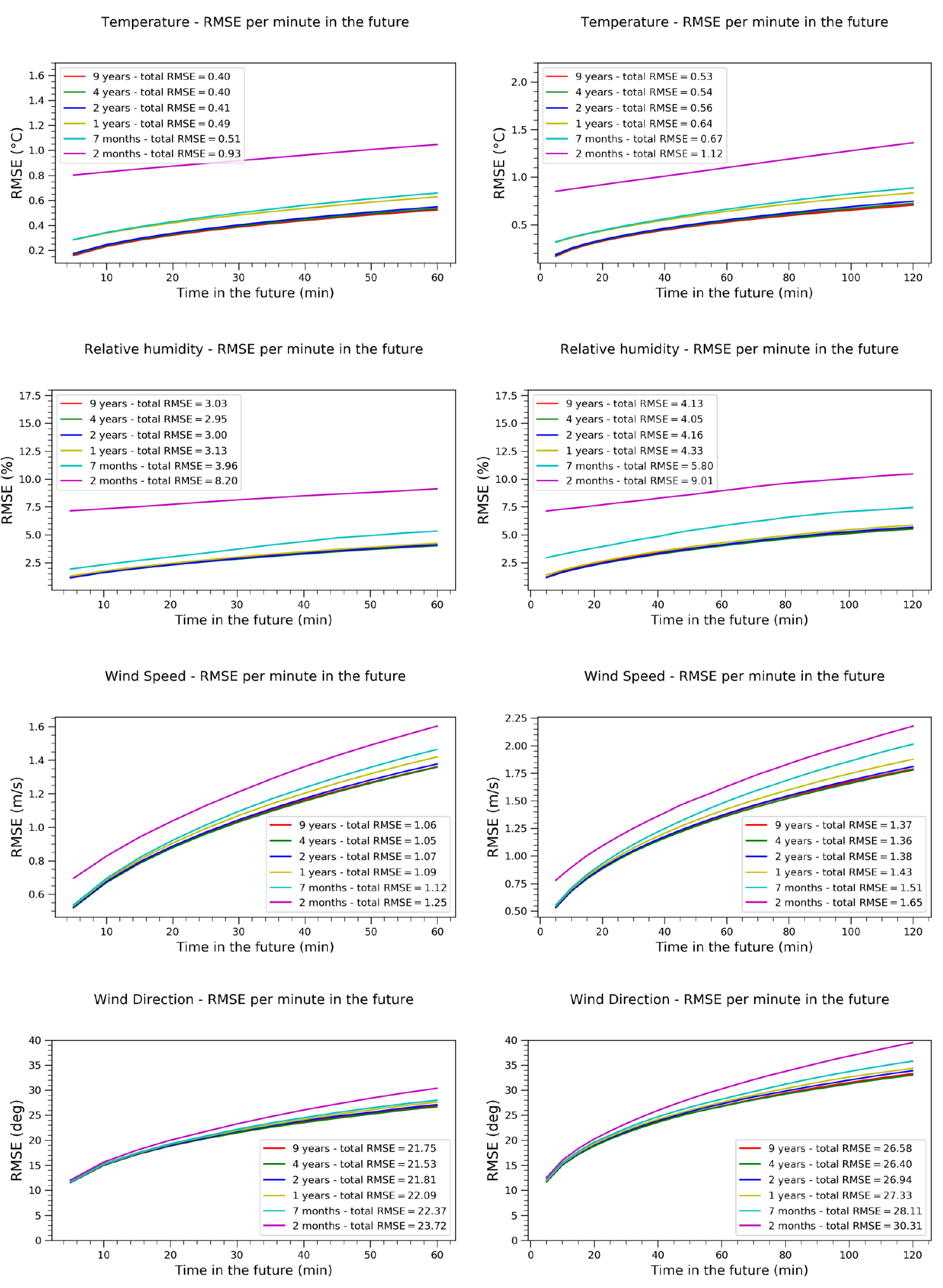}
\caption{RF prediction of atmospheric parameters with varying size of the training sets defined in table \ref{tab:trainlenatmo} (left). On the left column is the forecast performed for 1 hour in the future, on the right column the forecast relative to 2 hours in the future. From top to bottom row: Temperature at 30m from ground, Relative humidity at 30m from ground, Wind direction at 30m from ground, Wind speed at 30m from ground, Wind direction at 20m from ground. The value of the RMSE is the result of the average in the interval [0,1h] and [0h,2h], in table \ref{tab:rmsepartialatmo} the reader can find the partial RMSEs relative to 2 years long training.}
\label{fig:atmosc}
\end{figure}

\begin{figure}
\centering
\includegraphics[width=0.85\textwidth]{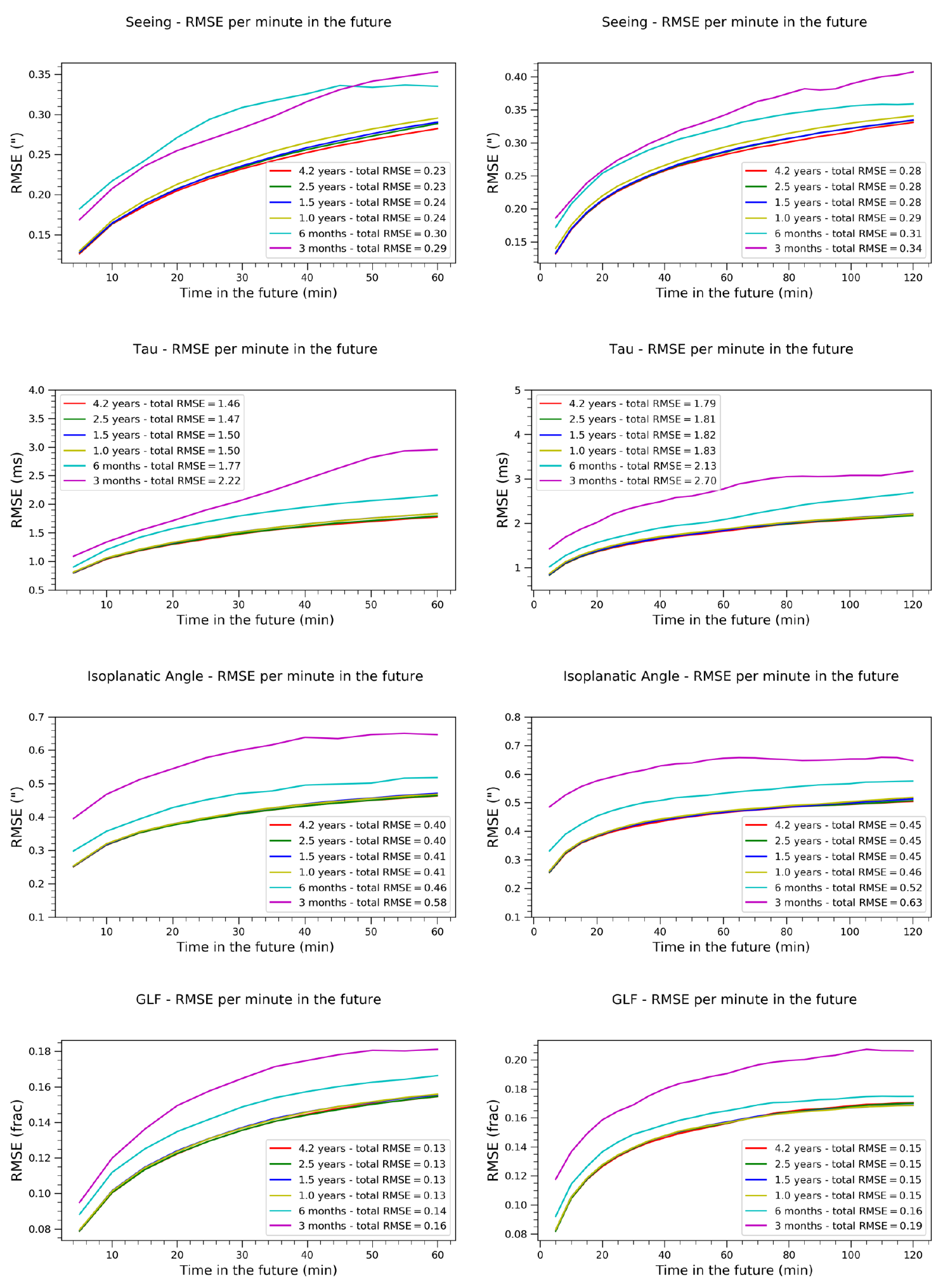}
\caption{RF prediction of astroclimatic parameters with varying size of the training sets defined in table \ref{tab:trainlenatmo} (right). On the left column is the forecast performed for 1 hour in the future, on the right column the forecast relative to 2 hours in the future. From top to bottom row: Seeing, $\tau_0$, $\theta_0$, GLF. The value of the RMSE is the result of the average in the interval [0,1h] and [0h,2h], in table \ref{tab:rmsepartialatmo} the reader can find the partial RMSEs relative to 1.5 years long training.}
\label{fig:astrosc}
\end{figure}

For atmospheric parameters we considered training data up to a maximum of 9 years included in the period [2010,2019], while the test set is 2.5 years included in the period [2019/01/01 - 2021/06/01] is the same for all the tested parameters. in Fig. \ref{fig:atmosc} we see that the performance saturation level is achieved after 1-2 years of training length, depending of the parameter. We conclude that, for atmospheric parameters, a training time of 2 years is sufficient to obtain performances at the saturation level either for the forecast at 1h (left column) and 2h (right column) in the future. \\
For OT astroclimatc parameters we considered training data up to a maximum of 4.2 years included in the period [2016,2030], while the test sample testing sample is taken equal to 0.5 year included in the period [2020/07/01 -
2020/12/31]. in Fig. \ref{fig:astrosc} we observe that the performance saturation level is achieved after 1.5 years of training data length.\\
In figures \ref{fig:atmosc} and \ref{fig:astrosc} we report the RMSEs computed as an average over 1h and 2h forecast respectively (left and right column), while in table \ref{tab:rmsepartialatmo} we report the partial RMSEs computed over the first and second hour of the 2-hours long forecast, relative to the saturation level performance training sets identified above (2 years and 1.5 years respectively for atmospheric and OT parameters).\\

%\begin{figure}
%\centering
%\includegraphics[width=1.\textwidth]{tempsc.pdf}
%\caption{Temperature at 30m from ground - RF prediction over one hour in the future (left) and 2 hours in the future (right). Time axis is in minutes. Results are obtained over different training set length defined in table \ref{tab:trainlenatmo}. }
%\label{fig:tempsc}
%\end{figure}
%\begin{figure}
%\centering
%\includegraphics[width=1.\textwidth]{rhsc.pdf}
%\caption{Relative humidity at 30m from ground - RF prediction over one hour in the future (left) and 2 hours in the future (right). Time axis is in minutes. Results are obtained over different training set length defined in table %\ref{tab:trainlenatmo}. }
%\label{fig:rhsc}
%\end{figure}
%\begin{figure}
%\centering
%\includegraphics[width=1.\textwidth]{wssc.pdf}
%\caption{Wind speed at 20m from ground - RF prediction over one hour in the future (left) and 2 hours in the future (right). Time axis is in minutes. Results are obtained over different training set length defined in table \ref{tab:trainlenatmo}. }
%\label{fig:wssc}
%\end{figure}
%\begin{figure}
%\centering
%\includegraphics[width=1.\textwidth]{wdsc.pdf}
%\caption{Wind direction at 20m from ground - RF prediction over one hour in the future (left) and 2 hours in the future (right). %Time axis is in minutes. Results are obtained over different training set length defined in table \ref{tab:trainlenatmo}. }
%\label{fig:wdsc}
%\end{figure}

\begin{table}[ht]
\caption{RSME computed for 1h forecasts and 2h forecasts time windows. In the latter case we separate the RMSE computed over the first and second hour. All values are relative to the 2 years training set used in Fig. \ref{fig:atmosc} (atmospheric parameters) and to 1.5 years training set used in Fig. \ref{fig:astrosc} (astroclimatic parameters), which correspond to the respective saturation levels.}
\label{tab:rmsepartialatmo}
\begin{center}       
\begin{tabular}{|l|c|c|c|} %% this creates two columns
%% |l|l| to left justify each column entry
%% |c|c| to center each column entry
%% use of \rule[]{}{} below opens up each row
\hline
{\bf Parameter} & {\bf 1h forecast} & \multicolumn{2}{c}{ {\bf 2h forecast}} \\
\hline
{\bf } & {\bf 1h RMSE} & {\bf 1h RMSE} & {\bf 2h RMSE} \\
\hline
Temperature & 0.41°C & 0.42°C & 0.67°C \\
\hline
Relative Humidity & 3.0\% & 3.0\% & 5.1\% \\
\hline
Wind Speed & 1.07m/s & 1.07m/s & 1.64m/s \\
\hline
Wind Direction & 21.8° & 21.9° & 31.2° \\
\hline
Seeing & 0.24'' & 0.24'' & 0.32'' \\
\hline
$\tau_0$ & 1.5ms & 1.5ms & 2.0ms \\
\hline
$\theta_0$ & 0.41'' & 0.41'' & 0.49'' \\
\hline
GLF & 13 (\%)& 14 (\%) & 17 (\%) \\
\hline
\end{tabular}
\end{center}
\end{table}

\begin{figure}
\centering
\includegraphics[width=1.\textwidth]{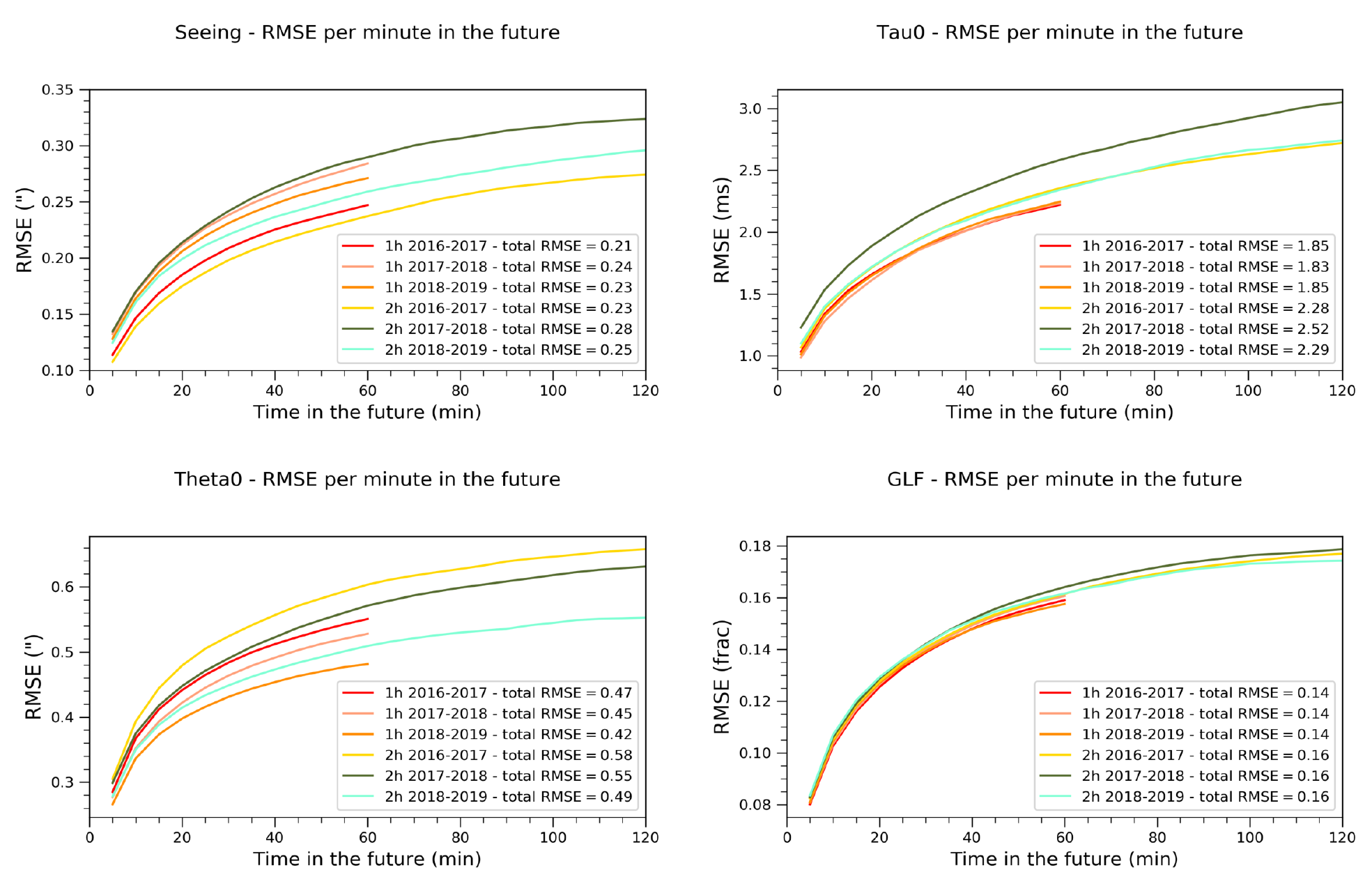}
\caption{Scaling of forecast performance, with respect to a change of training and test sample length. For both cases (1h and 2h), we selected 1.5 year for the training (see Fig. 2) and 0.5 year for the test. The test is performed on the OT parameters.}
\label{fig:diffset}
\end{figure}

A second test that we decided to carry on in this paper is to check how the forecast performance of RF is dependent on a different choice of training/test set combinations. This is fundamental to ensure that the performance is not tied to peculiar conditions present in either the training or test set chosen for the application. In Fig. \ref{fig:diffset} we report the result of this test performed only on the OT parameters, for the sake of simplicity. In all the tests we selected one year for the training and one year for the testing, changing the selection each time (e.g. 2017-2018 respectively). All tests were performed both for 1h and 2h future time windows and input data were resampled over 5 minutes.\\
This preliminary analysis indicates a variability of the RMSE that smaller at 1h than at 2h. From the results we observe that the order of magnitude of the maximum error bar (i.e. that at 2h) is: $\Delta\sim0.05''$ for the seeing, $\Delta\sim 0.24ms''$ for $\tau_0$, $\Delta\sim0.04''$ for $\theta_0$ and $\Delta\sim0.02$ for GLF. This results also gives us a characterization of the reliability and the intrinsic accuracy of the ML the method, since it tells how the results are dependent on the choice of the training and test sets.\\

\begin{figure}
\centering
\includegraphics[width=0.7\textwidth]{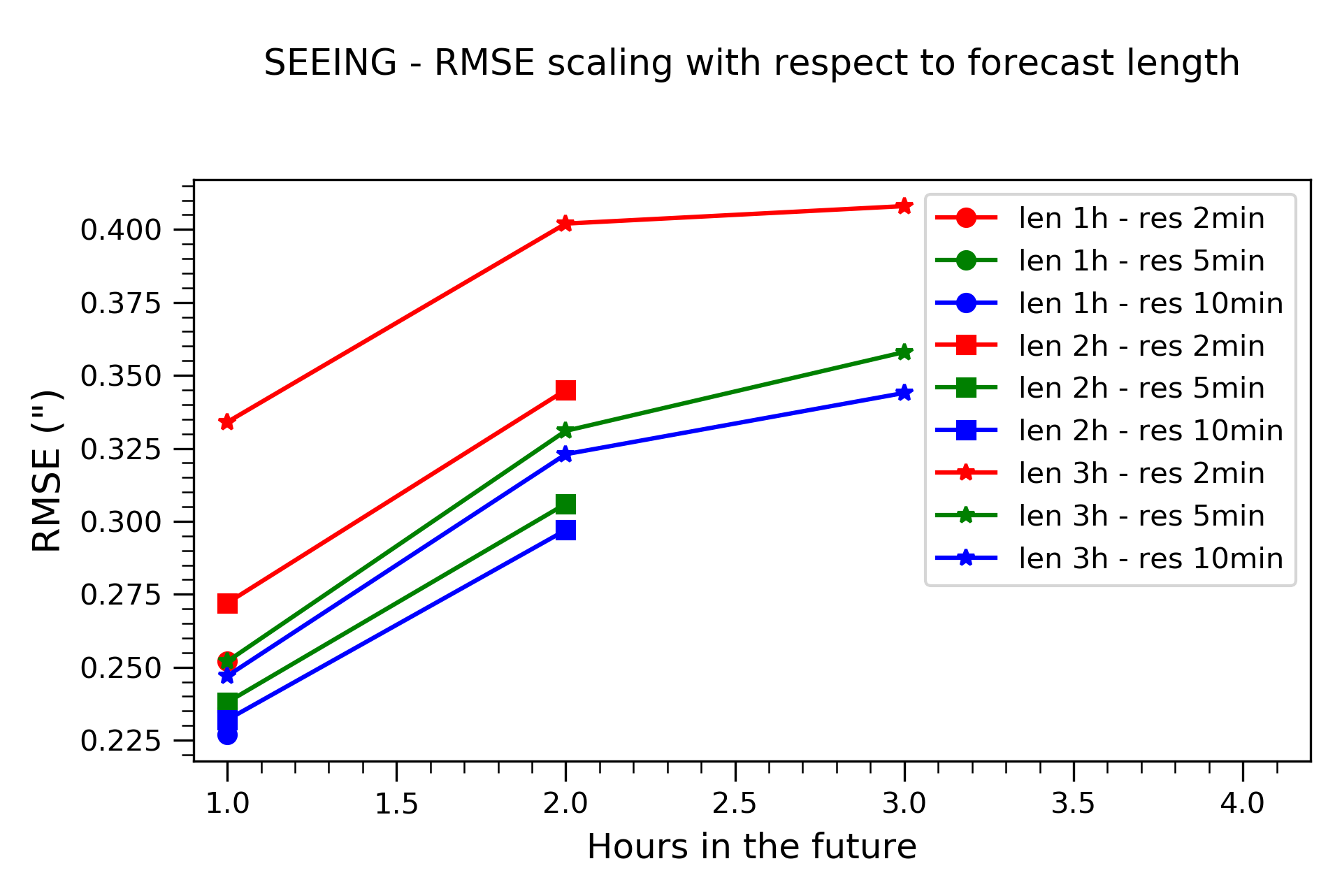}
\caption{Scaling of forecast performance, with respect to a change of resampling time. The test is performed only on the seeing parameter.}
\label{fig:scaleres}
\end{figure}

The final test performed in this contribution involves checking how the RF performances scale with the choice of the resampling time of the input data. The tests above used a 5 min resampling, while here we tested 2 min, 5 min and 10 min resampling to show the relative comparison. We decided to test only the seeing parameter, which is the most critical for telescope applications. The test is performed on the 1.5 training set defined in table \ref{tab:trainlenatmo} (right table) and on the same 0.5 years test set used before in Fig. \ref{fig:astrosc}. We decided to extend the test to three hours in the future to see if the performance shows any sign of saturation with an increased forecast time window. We couldn't perform a reliable 1 min resample test because the original sampling of the data sometimes have holes larger than 1 min, thus it would leave to many NaNs when performing the resampling average operation (averages over empty time windows). Since we need consecutive time windows to perform the forecast, the presence of so many holes would leave us with too little usable data to obtain a statistically reliable result.\\
In Fig. \ref{fig:scaleres} we show the results of this test. It is evident how the performances increase uniformly with the increasing of the resampling time, which corresponds to making coarse-grained predictions as an average over a wide time-window. In any case, the difference between 5 min and 10 min resampling is sufficiently small to consider 5 min a reasonable trade-off between fine-grained prediction and forecast performance.\\

\section{Conclusions}
\label{sec:end}
Forecast of atmospheric conditions and optical turbulence plays a critical role for the optimization and planning of telescope operations. In this short contribution we explored the forecast of atmospheric and optical turbulence parameters made with a Machine Learning algorithm, called Random Forest Regression, and data from the VLT telemetry. No input from physics such as atmospheric forecast models was used. Specifically we tested the variability of the ML method performance with respect to different tunable parameters, such as data resampling, training sample length and different training/test samples. This allowed us to evaluate the reliability of the method itself and to pave the way to further implementations which may make use of the data produced by the atmospheric forecast systems that are currently planned at the Very Large Telescope. Our preliminary analyses indicate that, for the astroclimatic parameters a training of 1.5 years is enough to saturate the peak forecast performances, in case of atmospheric parameters is preferable a sample of 2 years. With respect to the choice of training and test sets, the variability of the performance is slightly larger at 2h with respect to 1h time scale. On a time scale of 2h our preliminary analyses indicate for the seeing a $\Delta\sim 0.05''$, for the $\tau_0$ $\Delta\sim 0.24$ms, for $\theta_0$ $\Delta\sim 0.04''$. On a time scale of 1h values of $\Delta$ are even smaller than this. This means that the method is statistically robust and can be applied with a relative level of reproducibility. The forecast sampling is a critical choice, since a larger sampling of the forecasted data corresponds directly to better performance, so this parameter must be tuned to find a balance with respect to the requested performance level. Future evolution of our study will focus on different algorithms (such as Multi Layer Perceptron - MLP) and on a use of ML algorithms joint to atmospheric models.

\acknowledgments
Study co-funded by the FCRF foundation action - N. 45103, ALTA Center (ENV002), Horizon 2020, Research and Innovation, SOLARNET (N. 8241135).


\begin{thebibliography}{1}


\bibitem{turchi2020}
Turchi, A., Masciadri, E., Pathak, P, Kasper, M., 
``High-accuacry short-term precipiatable water-vapour operational forecast at the Very large telescope and perspectives for sky background'', {\em 497}, 4910, (2020)

\bibitem{rabien2019}
Rabien, S., Angel, R., Barl. L., et al., 
``ARGOS at the LBT'', {\em A\&A}, {\bf 621}, (2019)

\bibitem{neichel2014}
Neichel, B., Rigaut, F., Vidal, F. et al., 
``Gemini multiconjugate adaptive optics system review – II. Commissioning, operation and overall performance'', {\em MNRAS}, {\bf 440}, 1002, (2014)

\bibitem{pedichini2016}
Pedichini, F., Ambrosino, F., Centrone, M., et al., 
``The V-SHARK high contrast imager at LBT'', Proc. SPIE vol. 9908, id. 990832, (2016)

\bibitem{masciadri2020}
Masciadri, E., Martellioni, G, Turchi, A., 
``Filtering techniques to enhance optical turbulence forecast performances at short time-scales'', {\em MNRAS}, {\bf 492}, 140-152, (2020)

\bibitem{turchi2017}
Turchi, A., Masciadri, E., Fini. L.,
``Forecasting surface-layer atmospheric parameters at the Large Binocular Telescopes'', {\em MNRAS}, {\bf 466}, 1925-1943, (2017)

\bibitem{masciadri2022}
Masciadri, E., Turchi, A., Fini, L., ``Optical turbulence forecasts at short time scales at the VLTl'', {\em MNRAS}, submitted, (2022)

\bibitem{masciadri1999}
Masciadri, E., Vernin, J., Bougeault, P., ``3D mapping of optical turbulence using an atmospheric numerical model'', {\em A\&ASS}, {\bf 137}, 185, (1999)

\bibitem{milli2019}
Milli, J., Gonzalez, R., Fluxa, P.R., et al., 
``Nowcasting the turbulence at the Paranal Observatory'', proc. AO4ELT6, arXiv:1910.13767v1, (2019)

\bibitem{andy2002}
Liaw, A., Wiener, M.,
``Classification and Regression by randomForest'', {\em R News}, {\bf 2}, 18-22, (2002)

\bibitem{pedregosa2011}
Pedregosa, F., Varoquaux, G., Gramfort, A., et. al.,
``Scikit-learn: Machine Learning in {P}ython'', {\em Journal of Machine Learning Research}, {\bf 12}, 2825-2830, (2011)

\bibitem{masciadri2013}
Masciadri, E., Lascaux, F., Turchi, A., Fini, L.,
``MOSE: operational forecast of the optical turbulence and atmospheric parameters at the European Southern Observatory ground-based sites I. Overview and vertical stratification of the atmosphere'', {\em MNRAS}, {\bf 436}, 1968-1985, (2013)

\bibitem{sivo2018}
Sivo, G., Turchi, A., Masciadri, E., Guesalaga, A.,Neichel, B.,
``Towards an automatic wind speed and direction profiler for Wide Field adaptive optics systems '', {\em MNRAS}, {\bf 476}, 999-1009, (2018)

%\bibitem{murtagh1991}
%Murtagh, F., ``Multilayer perceptrons for classification and regression'', {\em Neurocomputing}, {\bf 2}, 183-197, (1991)


\end{thebibliography}
\end{document}